\documentclass[aps,prl,twocolumn,groupedaddress]{revtex4}

\usepackage{graphics}
\usepackage[T1]{fontenc}
\usepackage[latin9]{inputenc}
\usepackage{amsbsy}
\usepackage{graphicx}
\usepackage{amssymb}
\usepackage{amsmath}
\usepackage{wasysym}

\newcommand{\be}{\begin{equation}}
\newcommand{\ee}{\end{equation}}
\newcommand{\bel}[1]{\begin{equation}\label{#1}}
\newcommand{\bea}{\begin{eqnarray}}
\newcommand{\eea}{\end{eqnarray}}
\newcommand{\bef}{\begin{figure}}
\newcommand{\enf}{\end{figure}}
\newcommand{\ba}{\begin{array}}
\newcommand{\ball}{\begin{array}{ll}}
\newcommand{\bacl}{\begin{array}{cl}}
\newcommand{\bacll}{\begin{array}{cll}}
\newcommand{\bal}{\begin{array}{l}}
\newcommand{\bac}{\begin{array}{c}}
\newcommand{\ea}{\end{array}}

\renewcommand{\i}{\mathrm i}

\begin{document}

\bibliographystyle{apsrev}

\title{Small-world spectra in mean field theory}

\author{Carsten Grabow${}^{a}$}%
\author{Stefan Grosskinsky${}^{b}$}
\author{Marc Timme${}^{a,c}$}%

\affiliation{%
${}^a$Network Dynamics Group, Max Planck Institute for Dynamics and Self-Organization (MPIDS), 37073 G\"{o}ttingen, Germany\\
${}^b$Mathematics Institute and Centre for Complexity Science, University of Warwick, Coventry CV4 7AL, UK\\
${}^c$Bernstein Center for Computational Neuroscience G\"{o}ttingen, 37073
G\"{o}ttingen, Germany}%

\date{\today}

\begin{abstract}
Collective dynamics on small-world networks emerge in a broad range of systems with their spectra characterizing fundamental asymptotic features. Here we derive analytic mean field predictions for the spectra of small-world models that systematically interpolate between regular and random topologies by varying their randomness. These theoretical predictions agree well with the actual spectra (obtained by numerical diagonalization) for undirected and directed networks and from fully regular to strongly random topologies. These results may provide analytical insights to empirically found features of dynamics on small-world networks from various research fields, including biology, physics, engineering and social science.
\end{abstract}


\maketitle

Many networked systems, including the internet \cite{Broder:2000tf}, power grids \cite{Watts:1998vz}, airline traffic \cite{Amaral:2000wk}, polymers \cite{Jespersen:434532}, metabolic pathways \cite{Wagner:2001ga}, social networks \cite{Watts:1998vz} and neural circuits \cite{celegans} share two structural characteristics: a high 'clustering' such that two nodes connected to a joint third are likely to also be connected to each other; and a short average path length, meaning that the path that connects two randomly chosen nodes is short on average (known as the small-world effect). 

These topological features of small-world networks underly their collective dynamics such as synchronization, diffusion, relaxation and coordination processes \cite{SynchBook,Strogatz01}. In particular, the asymptotic dynamics on a small world is characterized by its graph Laplacian. Such processes occur in various fields ranging from opinion formation in social networks \cite{Pluchino05} and consensus dynamics of agents \cite{OlfatiSaber:2005wh} to synchronization in biological circuits \cite{Grabow:2010cl,Grabow:2011cu} and relaxation oscillations in gene regulatory networks \cite{McMillen:2002fw,Gardner03}.

In standard models of small-world topologies \cite{Watts:1998vz} a continuous parameter, the topological randomness $q\in [0,1]$, interpolates between fully regular ($q=0$), small-world (low $q \ll1$) and fully random networks ($q = 1$). Typically, networks for $q>0$ are constructed by randomly rewiring a fraction $q$ of edges. Although such models based on rewiring have received massive attention both theoretically and in applications (as certified, e.g., by the huge number of references to the original work \cite{Watts:1998vz}), for most of their features analytical predictions are not known to date (for a mean field solution of the average path length see Ref. \cite{Newman:2000vd}). In particular, the spectrum of small-world Laplacians has been studied for several specific cases and numerically \cite{Monasson:1999vo,Jost01,Barahona:2002bm,Mori04,Kuhn11}, but a general derivation of reliable analytic predictions is still missing. 

In this Letter we present an analytic derivation of small-world spectra based on a two-stage mean field approximation that we introduce. A single formula covers the entire spectrum from regular via small-world to strongly randomized topologies, explaining also the simultaneous dependencies on network size, average degree and randomness $q$. Numerical diagonalization of Laplacians of undirected and directed networks shows that the analytic prediction well approximates all actual eigenvalues, except for extremal parameter settings such as $q$ of the order of unity, where standard random matrix theory can be applied.

\begin{figure}
\includegraphics[width=85mm]{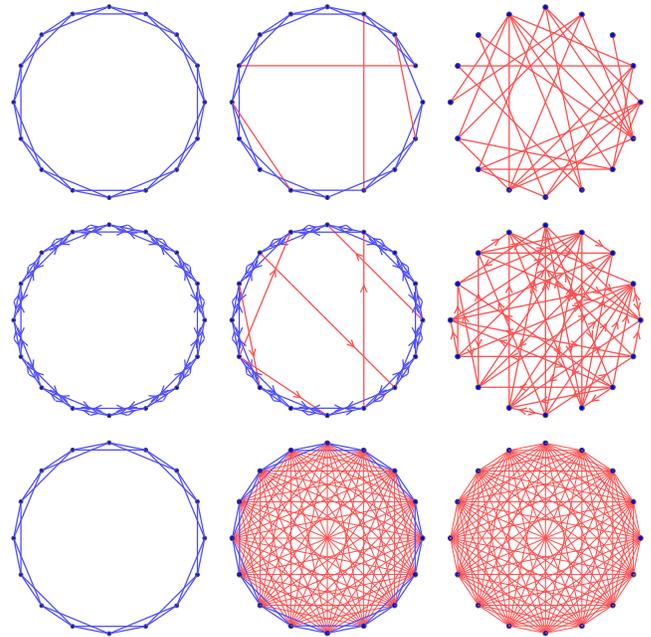}
\caption{\label{fig:networks} (color online) Rewiring 'on average' (Cartoon for $N=16$ and $k=4$). Single realizations of rewiring for (a) undirected and (b) directed networks; (c) mean field rewiring.  From left to right: $q=0$
(regular ring network), $q=0.1$ ('small world') and $q=1$ (random network). The regular ring network is the same for (a), (b) and (c).}
\end{figure}

Consider a graph of $N$ nodes on a one-dimensional ring lattice with periodic boundary conditions. Each node receives links from its $k/2$ nearest neighbors on both sides (for simplicity of presentation we take $k$ and $N$ to be even). Randomness is introduced by rewiring. Following \cite{Watts:1998vz} for undirected networks, we first cut each edge with probability $q$. Afterwards all cut edges are rewired to nodes chosen uniformly at random from the whole network (avoiding double edges and self-loops). Similarly, for directed networks \cite{Fagiolo:2007th}, we first cut all tails of edges with probability $q$ and rewire them afterwards (avoiding double edges and self-loops as well). 

The generic asymptotic relaxation dynamics on such a network is characterized by its graph Laplacian $\tilde{\Lambda}$ defined by its elements
\begin{equation}
\tilde{\Lambda}_{ij}=A_{ij}(1-\delta_{ij}) - k_{i} \delta_{ij}  \label{eq:laplacian}
\end{equation}
\noindent for $i,j \in \{1,\dots,N\} $, where $A_{ij}$ are the elements of the adjacency matrix (one for an existing edge and zero for no edge), $k_{i}$ is the degree of node $i$ (replaced by the in-degree for directed networks) and $\delta_{ij}$ is the Kronecker-delta. Note that the spectrum of the Laplacian for a directed network is complex while the spectrum for an undirected one is real. 

What is the spectrum of these networks in dependence on the network size $N$, the average degree $k$ and the topological randomness $q$? To analytically predict it, we introduce an average rewiring process (as depicted in Fig.~\ref{fig:networks} in comparison to both other rewiring procedures for undirected and directed networks): Define a circulant mean field Laplacian
\begin{equation}
\tilde{\Lambda}^{\textrm{mf}}=\begin{pmatrix}  c_{0}	  & c_{1} & c_{2}    &               & \cdots & c_{N-1} \\
			      c_{N-1} & c_{0} & c_{1}   & c_{2}     &  		& \vdots  \\ 
			      		  & c_{N-1} & c_{0} & c_{1}     & \ddots & 		 \\
			      \vdots   & 	          & \ddots & \ddots   & \ddots & c_{2} \\
			      		  &  &  &         &  & c_{1} \\
				c_{1} &  \cdots &  &         & c_{N-1} & c_{0}   \end{pmatrix} \ .
				\label{eq:mftlaplacian}
\end{equation}
For the initial ring ($q=0$), Eq.~(\ref{eq:mftlaplacian}) is exact and the matrix elements take the form
\begin{equation}
c_{i}=\left\{ \begin{array}{ll}
-k & \mbox{if}\  i = 0 \\
1 & \mbox{if}\  i \in \{1,\hdots,\frac{k}{2},N-\frac{k}{2},\hdots,N-1\}=:S_{1} \\
0 & \mbox{if}\ i \in \{\frac{k}{2}+1,\hdots, N-\frac{k}{2}-1\}=:S_{2} \ , \end{array}\right .  \label{eq:ringmatrixelements}
\end{equation}
where we classify the elements into those representing the original ring $S_{1}$ and those representing absent edges $S_{2}$ outside that ring. 

For given $q>0$, instead of rewiring each edge independently with a certain probability to obtain a specific randomized network, we now 'rewire on average' to obtain a mean field version of the randomized network ensemble: Firstly, the average total weight $q k N/2$ representing all edges to be rewired is subtracted uniformly from the weights of existing edges of $S_{1}$. Secondly, the rewired weight is distributed uniformly among the total 'available' weight in the whole network. The latter is given by $f=[N(N-1)-(1-q) k N]/2$, where each edge is assumed to carry at most weight one. Of this total, the weight $f_{1}=q k N/2$ is available in $S_{1}$ and  $f_{2}=[N(N-1)- k N]/2$ in $S_{2}$. The fraction $f_{1}/f$ is then assigned to elements representing edges in $S_{1}$ and the fraction $f_{2}/f$ to those representing $S_{2}$. Therefore, an individual edge in $S_{1}$ gets the additional weight 
\be
w_{1}= \frac{f_{1}}{f} \frac{\frac{qkN}{2}}{\frac{kN}{2}}=\frac{q^{2}k}{N-1-(1-q)k} \ ,
\ee
and an edge in $S_{2}$ gets the new weight
\be
w_{2}= \frac{f_{2}}{f} \frac{\frac{qkN}{2}}{\frac{N(N-1)-kN}{2}}=\frac{qk}{N-1-(1-q)k} \ .
\ee
Thus, in our mean field theory the elements of the Laplacian (\ref{eq:mftlaplacian}) of a network on $N$ nodes with degree $k$ after rewiring with probability $q$ are given by
\begin{equation}
c_{i}=\left\{ \begin{array}{ll}
-k & \mbox{if}\  i=0 \\
1-q+ w_{1} & \mbox{if}\   i \in S_{1} \\
w_{2} & \mbox{if}\ i \in S_{2} \ .\end{array}\right.\label{eq:eigenmatrixelements}
\end{equation}
The mean field Laplacian defined by (\ref{eq:mftlaplacian}) and (\ref{eq:eigenmatrixelements}) by construction is a circulant matrix with eigenvalues \cite{circbook}
\begin{equation}
\tilde{\lambda}^{\textrm{mf}}_{l}=\sum^{N-1}_{j=0}c_{j} \exp\left({\frac{-2\pi \i (l-1) j}{N}}\right) \ .
\end{equation}
The trivial eigenvalue $\tilde{\lambda}^{\textrm{mf}}_{1} =0$ follows immediately. It is common to all networks (for all $q$, $N$, and any $k\leq N-1$) and reflects the invariance of Laplacian dynamics against uniform shifts, as seen from the associated eigenvector $\tilde{v}_1=(1,\ldots,1)^{\textsf{T}}$. 
Exploiting the additional transposition symmetry $c_{j}=c_{N-j}$
yields an analytical expression for the remaining spectrum ($l \in \{2,\dots,N\}$), 
\begin{align}
\tilde{\lambda}^{\textrm{mf}}_{l}(N,k,q)=  & -k +  k c^\prime \sum^{\frac{k}{2}}_{j=1} x_{l}^{j}  \nonumber \\
&+ k c^{\prime\prime} \sum^{N-1-\frac{k}{2}}_{j=\frac{k}{2}+1} x_{l}^{j} 
+  k c^\prime \sum^{N-1}_{j=N-\frac{k}{2}} x_{l}^{j}  \nonumber \\
=  & -k - k c^\prime + k \left(c^\prime -  c^{\prime\prime}\right) \frac{\sin \left(\frac{(k+1)(l-1) \pi}{N}\right)}{\left(\sin \frac{(l-1)\pi}{N}\right)} \ , \label{eq:mftspectrum}
\end{align}
where $x_{l}=\exp{[-2\pi \i (l-1)/N]}$, $c^\prime=(1-q)/k+ q c^{\prime\prime}$ and $c^{\prime\prime}=q/[N-1-(1-q)k]$.

As the offset of each eigenvalue (\ref{eq:mftspectrum}) equals $k$, we consider the scaled eigenvalues $\lambda^{\textrm{mf}}_{l}(N,k,q)=\tilde{\lambda}^{\textrm{mf}}_{l}(N,k,q)/k$ in the following to allow for a consistent analysis for different $k$.

We first focus on the second largest eigenvalue since it dominates the long time dynamics (see e.g. \cite{Grabow:2010cl,Grabow:2011cu}). Monotonicity considerations show that the second largest eigenvalue is given for $l=2$ where eq.~(\ref{eq:mftspectrum}) simplifies to
\bel{eq:mft}
\lambda^{\textrm{mf}}_{2}(N,k,q)= -1 - c^\prime + \left(c^\prime -  c^{\prime\prime}\right) \frac{\sin \left(\frac{(k+1)\pi}{N}\right)}{\sin \left(\frac{\pi}{N}\right)} .
\ee

\begin{figure}
\includegraphics[width=85mm]{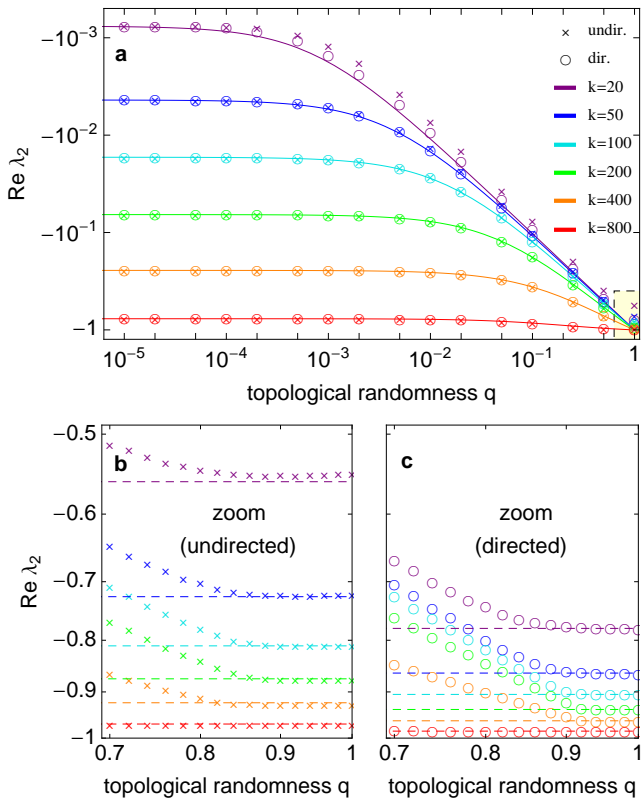}
\caption{(color online) Accuracy of analytic prediction of second largest eigenvalues as a function of topological randomness. (a) Numerical measurements for undirected  ($\times$) and directed ($\ocircle$) networks in comparison with the analytical mean field predictions (Eq.~(\ref{eq:mft}), solid lines) as a function of $q$, for different degrees $k$. (b) Zoom close to $q=1$ for undirected networks with the analytical predictions $\lambda^{\textrm{wsc}}_{2}$ via Wigner's semi-circle law (dashed lines). (c) Zoom close to $q=1$ for directed networks with the analytical predictions $\lambda^{\textrm{rmt}}_{2}$ from the theory of asymmetric random matrices (dashed lines). In (a), (b) and (c) error bars on the numerical measurements are smaller than the data points ($N=1000$, each data point averaged over $100$ realizations).
\label{fig:lambda2vsq}}
\end{figure}

The analytic prediction (\ref{eq:mft}) fits well with the eigenvalues of small worlds obtained by actual rewiring, cf.~Fig.~\ref{fig:lambda2vsq}. It turns out that the analytic prediction is accurate for both undirected and directed networks, and for all but very small relative degrees $k/N$. For small $k$, the prediction shows some deviation from the numerical results, but still is a good guide for the general dependence of the second largest eigenvalue on $q$. Moreover, the prediction (\ref{eq:mft}) approximates well the actual dependence of $\lambda_2$ for all but large $q$ of the order of one, thus including regular rings, small-world and even more substantially randomized network topologies. Close to $q=1$ (Fig.~\ref{fig:lambda2vsq},b,c), the second largest eigenvalues for undirected networks are well predicted via random matrix theory by Wigner's semi-circle law \cite{wigner51,mehta} $\lambda^{\textrm{wsc}}_{2}(N,k,1)=2 \sqrt{1/k - 1/N} - 1$, the real parts for directed networks by the circular law \cite{Sommers:1988uq,Timme:2004tx,Timme:2006hl,Goetze2010} $\lambda^{\textrm{rmt}}_{2}(N,k,1)=\sqrt{1/k - 1/N} - 1$.

\begin{figure}
\includegraphics[width=85mm]{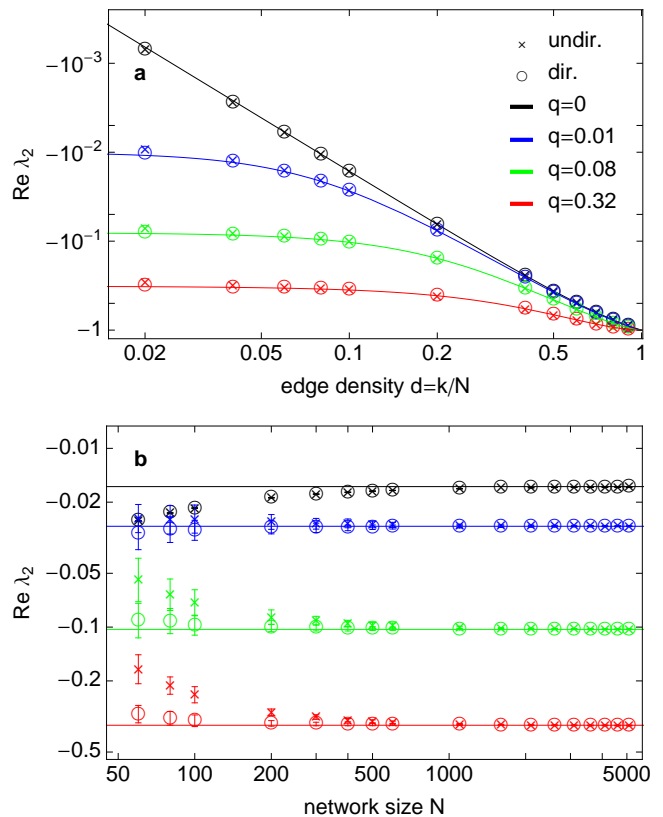}
\caption{(color online) Second largest eigenvalues in dependence on edge density and network size. (a) Numerical measurements for undirected ($\times$) and directed ($\ocircle$) networks in comparison with the analytic mean field prediction (\ref{eq:dfixed}). Error bars on the numerical measurements are smaller than the data points ($N=2000$, each data point averaged over $100$ realizations).
(b) Asymptotic ($N\rightarrow \infty$) real parts of the
second largest eigenvalues $\lambda_{2}$ in dependence on the network size $N$ for fixed edge density $d=k/N=0.1$ (each data point averaged over $100$ realizations; $q$-values and symbols as in (a)). \label{fig:lambda2dfixed}
}
\end{figure}

How does the second largest eigenvalue scale with the network size? Fixing the edge density $d=k/N$ for large $N \gg 1$ (ensuring that networks stay connected) yields the prediction
\begin{equation}
\lambda^{\textrm{mf}}_{2}(d,q) \simeq  -1 + \frac{(1-d)(1-q)}{(1-d (1-q))d \pi}\sin \left(d \pi \right) \label{eq:dfixed} \ 
\end{equation}
in the limit $N \rightarrow \infty$.
Our analytic prediction (\ref{eq:dfixed}) again approximates well the real part of the second largest eigenvalue in dependence on the edge density $d$ for networks of size above about $N=500$ nodes, for both undirected and directed networks, cf. Fig.~\ref{fig:lambda2dfixed}. For densities other than that displayed ($d=k/N=0.1$, Fig.~\ref{fig:lambda2dfixed} b) the real parts of the second largest eigenvalue show qualitatively the same asymptotic behaviour.

\begin{figure}
\includegraphics[width=85mm]{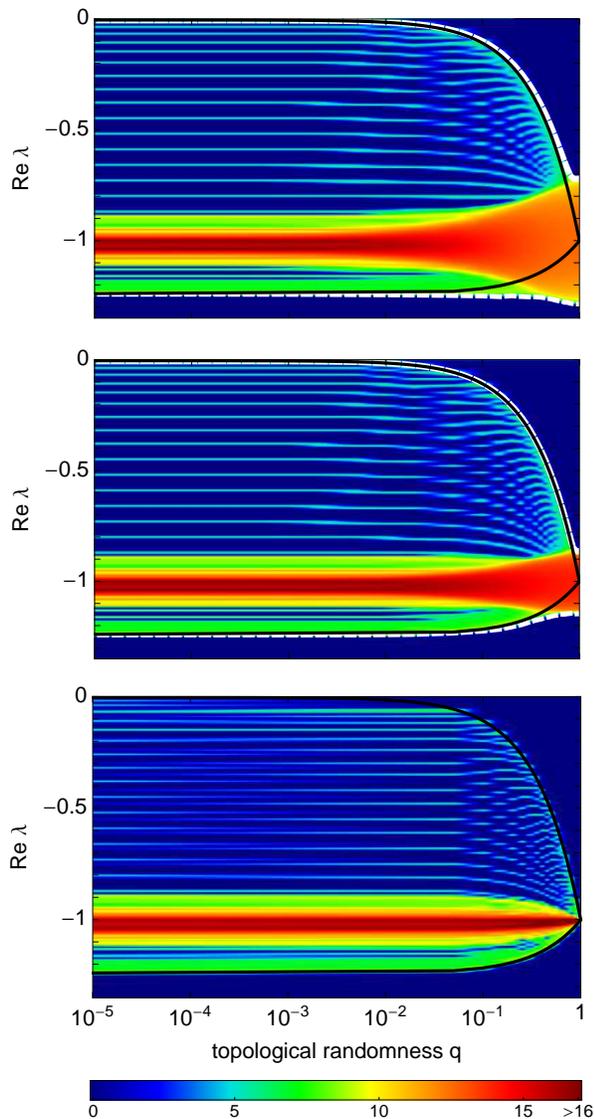} 
\caption{{(color online) Analytics predicts structure of entire spectrum. Densities of states, eq.~(\ref{eq:density}), for undirected (a), directed (b) and mean field (c) networks ($N=1000, k=50$). Dashed white lines show the extreme eigenvalues obtained numerically for undirected and directed networks. The solid black lines show the mean field prediction for the extreme eigenvalues. Densities of states for directed and undirected networks are averaged over $100$ realizations for a fixed $q$-value, while the mean field density is analytically determined by eq.~(\ref{eq:mftspectrum}). \label{fig:dos}}}
\end{figure}

To gain further insight into the entire spectrum we study the density of states $\rho (\lambda)$ (cf. e.g. \cite{Farkas:2001ig}) as defined in its discrete form, i.e. for finite network size $N$, by
\begin{equation}
\rho( \lambda ) =\frac{1}{N}\sum _{j=1}^{N} \delta(\lambda-\lambda_{j}) \ ,
\label{eq:density}
\end{equation}
where $\delta$ is the Dirac delta distribution. The evaluation of (\ref{eq:density}) for the analytic mean field predictions and for the numerically obtained eigenvalues of undirected and directed networks shows good qualitative agreement, cf.~Fig.~\ref{fig:dos}, for all but large topological randomness $q\rightarrow 1$. Spectra for networks with parameters other than $N=1000$ and $k=50$ yield qualitatively the same structure. Thus, the largest and smallest eigenvalues, the location and form of the bulk peaks as well as the entire structure of eigenvalues are well approximated (again, except for $q$ of order one) by the mean field predictions derived analytically.

In summary, we have introduced a simple two-stage mean field rewiring to analytically derive predictions for the spectra of graph Laplacians. Systematic numerical checks confirm that this prediction is accurate for the second largest eigenvalue for all parameter values except for small degrees or too large topological randomness of the order of unity. For smaller $k$, our analytic prediction still serves as a valuable guide for the overall dependence all topological parameters. For $q$ close to unity, our mean field predictions are well complemented by standard random matrix theory. Besides the second largest and smallest eigenvalues that already give valuable information about initial and asymptotic relaxation dynamics, also the bulk spectrum and the fine structure of the spectrum are well approximated by our analytical prediction. 

The spectral predictions in particular include regular rings, small-worlds, and substantially more randomly rewired networks and undirected as well as directed ones. Thus, our theoretical predictions agree well with the actual eigenvalues obtained numerically over almost the whole range of topological randomness $q$, thereby completing previous approaches based on perturbation theory for $q \ll 1$ \cite{Monasson:1999vo,Barahona:2002bm}. 

The approach presented here enables systematic analytic predictions of relaxation and consensus dynamics on randomized networks \cite{Watts:1998vz} in dependence of their key parameters $N$, $k$, and $q$, it substantially reduces computational efforts, and may make some regimes of interest accessible for the first time. Our mean field approach may be extended to rewiring processes starting from other than ring-like structures, e.g. to two or three dimensions, as for instance relevant for neural network modeling \cite{Sporns2009}. Checking with appropriate models, it may thus serve as a powerful tool to predict or deduce the relations between structural and dynamical properties of randomized networks. 

Supported by the BMBF, grant \# 01GQ1005B [MT], by the GGNB (DFG Grant \# GSC 226/1) [CG], by a grant of the
Max Planck Society [MT] and by an EPSRC grant \# EP/E501311/1 [SG].


\begin{thebibliography}{30}
\expandafter\ifx\csname natexlab\endcsname\relax\def\natexlab#1{#1}\fi
\expandafter\ifx\csname bibnamefont\endcsname\relax
  \def\bibnamefont#1{#1}\fi
\expandafter\ifx\csname bibfnamefont\endcsname\relax
  \def\bibfnamefont#1{#1}\fi
\expandafter\ifx\csname citenamefont\endcsname\relax
  \def\citenamefont#1{#1}\fi
\expandafter\ifx\csname url\endcsname\relax
  \def\url#1{\texttt{#1}}\fi
\expandafter\ifx\csname urlprefix\endcsname\relax\def\urlprefix{URL }\fi
\providecommand{\bibinfo}[2]{#2}
\providecommand{\eprint}[2][]{\url{#2}}

\bibitem[{\citenamefont{Broder et~al.}(2000)\citenamefont{Broder, Kumar,
  Maghoul, and Raghavan}}]{Broder:2000tf}
\bibinfo{author}{\bibfnamefont{A.}~\bibnamefont{Broder}},
  \bibinfo{author}{\bibfnamefont{R.}~\bibnamefont{Kumar}},
  \bibinfo{author}{\bibfnamefont{F.}~\bibnamefont{Maghoul}}, \bibnamefont{and}
  \bibinfo{author}{\bibfnamefont{P.}~\bibnamefont{Raghavan}},
  \bibinfo{journal}{Comput. Netw.} \textbf{\bibinfo{volume}{32}}
  (\bibinfo{year}{2000}).

\bibitem[{\citenamefont{Watts and Strogatz}(1998)}]{Watts:1998vz}
\bibinfo{author}{\bibfnamefont{D.}~\bibnamefont{Watts}} \bibnamefont{and}
  \bibinfo{author}{\bibfnamefont{S.}~\bibnamefont{Strogatz}},
  \bibinfo{journal}{Nature} \textbf{\bibinfo{volume}{393}},
  \bibinfo{pages}{440} (\bibinfo{year}{1998}).

\bibitem[{\citenamefont{Amaral et~al.}(2000)\citenamefont{Amaral, Scala,
  Barthelemy, and Stanley}}]{Amaral:2000wk}
\bibinfo{author}{\bibfnamefont{L.}~\bibnamefont{Amaral}},
  \bibinfo{author}{\bibfnamefont{A.}~\bibnamefont{Scala}},
  \bibinfo{author}{\bibfnamefont{M.}~\bibnamefont{Barthelemy}},
  \bibnamefont{and} \bibinfo{author}{\bibfnamefont{H.}~\bibnamefont{Stanley}},
  \bibinfo{journal}{Proc. Natl. Acad. Sci. U.S.A.}
  \textbf{\bibinfo{volume}{97}}, \bibinfo{pages}{11149} (\bibinfo{year}{2000}).

\bibitem[{\citenamefont{Jespersen et~al.}(2000)\citenamefont{Jespersen,
  Sokolov, and Blumen}}]{Jespersen:434532}
\bibinfo{author}{\bibfnamefont{S.}~\bibnamefont{Jespersen}},
  \bibinfo{author}{\bibfnamefont{I.~M.} \bibnamefont{Sokolov}},
  \bibnamefont{and} \bibinfo{author}{\bibfnamefont{A.}~\bibnamefont{Blumen}},
  \bibinfo{journal}{J. Chem. Phys.} \textbf{\bibinfo{volume}{113}}
  (\bibinfo{year}{2000}).

\bibitem[{\citenamefont{Wagner and Fell}(2001)}]{Wagner:2001ga}
\bibinfo{author}{\bibfnamefont{A.}~\bibnamefont{Wagner}} \bibnamefont{and}
  \bibinfo{author}{\bibfnamefont{D.~A.} \bibnamefont{Fell}},
  \bibinfo{journal}{Proc. R. Soc. London, Ser. B}
  \textbf{\bibinfo{volume}{268}}, \bibinfo{pages}{1803} (\bibinfo{year}{2001}).

\bibitem[{\citenamefont{Achacoso and Yamamoto}(1992)}]{celegans}
\bibinfo{author}{\bibfnamefont{T.}~\bibnamefont{Achacoso}} \bibnamefont{and}
  \bibinfo{author}{\bibfnamefont{W.}~\bibnamefont{Yamamoto}},
  \emph{\bibinfo{title}{{AY's Neuroanatomy of C. Elegans for Computation}}}
  (\bibinfo{publisher}{CRC Press, Boca Raton, FL}, \bibinfo{year}{1992}).

\bibitem[{\citenamefont{Pikovsky et~al.}(2001)\citenamefont{Pikovsky,
  Rosenblum, and Kurths}}]{SynchBook}
\bibinfo{author}{\bibfnamefont{A.}~\bibnamefont{Pikovsky}},
  \bibinfo{author}{\bibfnamefont{M.}~\bibnamefont{Rosenblum}},
  \bibnamefont{and} \bibinfo{author}{\bibfnamefont{J.}~\bibnamefont{Kurths}},
  \emph{\bibinfo{title}{Synchronization, A universal concept in nonlinear
  sciences}}, vol.~\bibinfo{volume}{12} of \emph{\bibinfo{series}{Cambridge
  Nonlinear Science Series}} (\bibinfo{publisher}{Cambridge University Press,
  Cambridge, UK}, \bibinfo{year}{2001}).

\bibitem[{\citenamefont{Strogatz}(2001)}]{Strogatz01}
\bibinfo{author}{\bibfnamefont{S.}~\bibnamefont{Strogatz}},
  \bibinfo{journal}{Nature} \textbf{\bibinfo{volume}{410}},
  \bibinfo{pages}{268} (\bibinfo{year}{2001}).

\bibitem[{\citenamefont{Pluchino et~al.}(2005)\citenamefont{Pluchino, Latora,
  and Rapisarda}}]{Pluchino05}
\bibinfo{author}{\bibfnamefont{A.}~\bibnamefont{Pluchino}},
  \bibinfo{author}{\bibfnamefont{V.}~\bibnamefont{Latora}}, \bibnamefont{and}
  \bibinfo{author}{\bibfnamefont{A.}~\bibnamefont{Rapisarda}},
  \bibinfo{journal}{Int. J. Mod. Phys. C} \textbf{\bibinfo{volume}{16}},
  \bibinfo{pages}{515} (\bibinfo{year}{2005}).

\bibitem[{\citenamefont{Olfati-Saber}(2005)}]{OlfatiSaber:2005wh}
\bibinfo{author}{\bibfnamefont{R.}~\bibnamefont{Olfati-Saber}},
  \bibinfo{journal}{Proc. Am. Control Conf.} \textbf{\bibinfo{volume}{4}},
  \bibinfo{pages}{2371} (\bibinfo{year}{2005}).

\bibitem[{\citenamefont{Grabow et~al.}(2010)\citenamefont{Grabow, Hill,
  Grosskinsky, and Timme}}]{Grabow:2010cl}
\bibinfo{author}{\bibfnamefont{C.}~\bibnamefont{Grabow}},
  \bibinfo{author}{\bibfnamefont{S.~M.} \bibnamefont{Hill}},
  \bibinfo{author}{\bibfnamefont{S.}~\bibnamefont{Grosskinsky}},
  \bibnamefont{and} \bibinfo{author}{\bibfnamefont{M.}~\bibnamefont{Timme}},
  \bibinfo{journal}{Europhys. Lett.} \textbf{\bibinfo{volume}{90}},
  \bibinfo{pages}{48002} (\bibinfo{year}{2010}).

\bibitem[{\citenamefont{Grabow et~al.}(2011)\citenamefont{Grabow, Grosskinsky,
  and Timme}}]{Grabow:2011cu}
\bibinfo{author}{\bibfnamefont{C.}~\bibnamefont{Grabow}},
  \bibinfo{author}{\bibfnamefont{S.}~\bibnamefont{Grosskinsky}},
  \bibnamefont{and} \bibinfo{author}{\bibfnamefont{M.}~\bibnamefont{Timme}},
  \bibinfo{journal}{Eur. Phys. J. B}  (\bibinfo{year}{2011}).

\bibitem[{\citenamefont{McMillen et~al.}(2002)\citenamefont{McMillen, Kopell,
  Hasty, and Collins}}]{McMillen:2002fw}
\bibinfo{author}{\bibfnamefont{D.}~\bibnamefont{McMillen}},
  \bibinfo{author}{\bibfnamefont{N.}~\bibnamefont{Kopell}},
  \bibinfo{author}{\bibfnamefont{J.}~\bibnamefont{Hasty}}, \bibnamefont{and}
  \bibinfo{author}{\bibfnamefont{J.}~\bibnamefont{Collins}},
  \bibinfo{journal}{Proc. Natl. Acad. Sci. U.S.A.}
  \textbf{\bibinfo{volume}{99}}, \bibinfo{pages}{679} (\bibinfo{year}{2002}).

\bibitem[{\citenamefont{Gardner et~al.}(2003)\citenamefont{Gardner,
  di~Bernardo, Lorenz, and Collins}}]{Gardner03}
\bibinfo{author}{\bibfnamefont{T.~S.} \bibnamefont{Gardner}},
  \bibinfo{author}{\bibfnamefont{D.}~\bibnamefont{di~Bernardo}},
  \bibinfo{author}{\bibfnamefont{D.}~\bibnamefont{Lorenz}}, \bibnamefont{and}
  \bibinfo{author}{\bibfnamefont{J.~J.} \bibnamefont{Collins}},
  \bibinfo{journal}{Science} \textbf{\bibinfo{volume}{301}},
  \bibinfo{pages}{102} (\bibinfo{year}{2003}).

\bibitem[{\citenamefont{Newman et~al.}(2000)\citenamefont{Newman, Moore, and
  Watts}}]{Newman:2000vd}
\bibinfo{author}{\bibfnamefont{M.~E.~J.}~\bibnamefont{Newman}},
  \bibinfo{author}{\bibfnamefont{C.}~\bibnamefont{Moore}}, \bibnamefont{and}
  \bibinfo{author}{\bibfnamefont{D.J.}~\bibnamefont{Watts}},
  \bibinfo{journal}{Phys. Rev. Lett.} \textbf{\bibinfo{volume}{84}},
  \bibinfo{pages}{3201} (\bibinfo{year}{2000}).

\bibitem[{\citenamefont{Monasson}(1999)}]{Monasson:1999vo}
\bibinfo{author}{\bibfnamefont{R.}~\bibnamefont{Monasson}},
  \bibinfo{journal}{Eur. Phys. J. B} \textbf{\bibinfo{volume}{12}},
  \bibinfo{pages}{555} (\bibinfo{year}{1999}).

\bibitem[{\citenamefont{Jost and Joy}(2001)}]{Jost01}
\bibinfo{author}{\bibfnamefont{J.}~\bibnamefont{Jost}} \bibnamefont{and}
  \bibinfo{author}{\bibfnamefont{M.~P.}~\bibnamefont{Joy}},
  \bibinfo{journal}{Phys. Rev. E} \textbf{\bibinfo{volume}{65}}
  (\bibinfo{year}{2001}).

\bibitem[{\citenamefont{Barahona and Pecora}(2002)}]{Barahona:2002bm}
\bibinfo{author}{\bibfnamefont{M.}~\bibnamefont{Barahona}} \bibnamefont{and}
  \bibinfo{author}{\bibfnamefont{L.~M.} \bibnamefont{Pecora}},
  \bibinfo{journal}{Phys. Rev. Lett.} \textbf{\bibinfo{volume}{89}},
  \bibinfo{pages}{4} (\bibinfo{year}{2002}).

\bibitem[{\citenamefont{Mori and Odagaki}(2004)}]{Mori04}
\bibinfo{author}{\bibfnamefont{F.}~\bibnamefont{Mori}} \bibnamefont{and}
  \bibinfo{author}{\bibfnamefont{T.}~\bibnamefont{Odagaki}},
  \bibinfo{journal}{J. Phys. Soc. Jpn.} \textbf{\bibinfo{volume}{73}},
  \bibinfo{pages}{3294} (\bibinfo{year}{2004}).

\bibitem[{\citenamefont{K{\"u}hn and van Mourik}(2011)}]{Kuhn11}
\bibinfo{author}{\bibfnamefont{R.}~\bibnamefont{K{\"u}hn}} \bibnamefont{and}
  \bibinfo{author}{\bibfnamefont{J.}~\bibnamefont{van Mourik}},
  \bibinfo{journal}{J. Phys. A} \textbf{\bibinfo{volume}{44}},
  \bibinfo{pages}{165205} (\bibinfo{year}{2011}).

\bibitem[{\citenamefont{Fagiolo}(2007)}]{Fagiolo:2007th}
\bibinfo{author}{\bibfnamefont{G.}~\bibnamefont{Fagiolo}},
  \bibinfo{journal}{Phys. Rev. E} \textbf{\bibinfo{volume}{76}},
  \bibinfo{pages}{026107} (\bibinfo{year}{2007}).

\bibitem[{\citenamefont{Golub and Van~Loan}(1996)}]{circbook}
\bibinfo{author}{\bibfnamefont{G.}~\bibnamefont{Golub}} \bibnamefont{and}
  \bibinfo{author}{\bibfnamefont{C.}~\bibnamefont{Van~Loan}},
  \emph{\bibinfo{title}{{Matrix Computations (Johns Hopkins Studies in Math.
  Sci.)}}} (\bibinfo{year}{1996}).

\bibitem[{\citenamefont{Wigner}(1951)}]{wigner51}
\bibinfo{author}{\bibfnamefont{E.~P.} \bibnamefont{Wigner}},
  \bibinfo{journal}{Proc. Cambridge Philos. Soc.}
  \textbf{\bibinfo{volume}{47}}, \bibinfo{pages}{790} (\bibinfo{year}{1951}).

\bibitem[{\citenamefont{Mehta}(1991)}]{mehta}
\bibinfo{author}{\bibfnamefont{M.}~\bibnamefont{Mehta}},
  \emph{\bibinfo{title}{{Random Matrices (Academic Press, New York)}}}
  (\bibinfo{year}{1991}).

\bibitem[{\citenamefont{Sommers et~al.}(1988)\citenamefont{Sommers, Crisanti,
  Sompolinsky, and Stein}}]{Sommers:1988uq}
\bibinfo{author}{\bibfnamefont{H.~J.}~\bibnamefont{Sommers}},
  \bibinfo{author}{\bibfnamefont{A.}~\bibnamefont{Crisanti}},
  \bibinfo{author}{\bibfnamefont{H.}~\bibnamefont{Sompolinsky}},
  \bibnamefont{and} \bibinfo{author}{\bibfnamefont{Y.}~\bibnamefont{Stein}},
  \bibinfo{journal}{Phys. Rev. Lett.} \textbf{\bibinfo{volume}{60}},
  \bibinfo{pages}{1895} (\bibinfo{year}{1988}).

\bibitem[{\citenamefont{Timme et~al.}(2004)\citenamefont{Timme, Wolf, and
  Geisel}}]{Timme:2004tx}
\bibinfo{author}{\bibfnamefont{M.}~\bibnamefont{Timme}},
  \bibinfo{author}{\bibfnamefont{F.}~\bibnamefont{Wolf}}, \bibnamefont{and}
  \bibinfo{author}{\bibfnamefont{T.}~\bibnamefont{Geisel}},
  \bibinfo{journal}{Phys. Rev. Lett.} \textbf{\bibinfo{volume}{92}},
  \bibinfo{pages}{074101} (\bibinfo{year}{2004}).

\bibitem[{\citenamefont{Timme et~al.}(2006)\citenamefont{Timme, Geisel, and
  Wolf}}]{Timme:2006hl}
\bibinfo{author}{\bibfnamefont{M.}~\bibnamefont{Timme}},
  \bibinfo{author}{\bibfnamefont{T.}~\bibnamefont{Geisel}}, \bibnamefont{and}
  \bibinfo{author}{\bibfnamefont{F.}~\bibnamefont{Wolf}},
  \bibinfo{journal}{Chaos} \textbf{\bibinfo{volume}{16}},
  \bibinfo{pages}{015108} (\bibinfo{year}{2006}).

\bibitem[{\citenamefont{G{\"o}tze and Tikhomirov}(2010)}]{Goetze2010}
\bibinfo{author}{\bibfnamefont{F.}~\bibnamefont{G{\"o}tze}} \bibnamefont{and}
  \bibinfo{author}{\bibfnamefont{A.}~\bibnamefont{Tikhomirov}},
  \bibinfo{journal}{Ann. Probab.} \textbf{\bibinfo{volume}{38}},
  \bibinfo{pages}{1444} (\bibinfo{year}{2010}).

\bibitem[{\citenamefont{Farkas et~al.}(2001)\citenamefont{Farkas, Der{\'e}nyi,
  Barab{\'a}si, and Vicsek}}]{Farkas:2001ig}
\bibinfo{author}{\bibfnamefont{I.~J.} \bibnamefont{Farkas}},
  \bibinfo{author}{\bibfnamefont{I.}~\bibnamefont{Der{\'e}nyi}},
  \bibinfo{author}{\bibfnamefont{A.-L.} \bibnamefont{Barab{\'a}si}},
  \bibnamefont{and} \bibinfo{author}{\bibfnamefont{T.}~\bibnamefont{Vicsek}},
  \bibinfo{journal}{Phys. Rev. E} \textbf{\bibinfo{volume}{64}},
  \bibinfo{pages}{12} (\bibinfo{year}{2001}).

\bibitem[{\citenamefont{Sporns and Bullmore}(2009)}]{Sporns2009}
\bibinfo{author}{\bibfnamefont{O.}~\bibnamefont{Sporns}} \bibnamefont{and}
  \bibinfo{author}{\bibfnamefont{E.}~\bibnamefont{Bullmore}},
  \bibinfo{journal}{Nat. Rev. Neurosci.} \textbf{\bibinfo{volume}{10}},
  \bibinfo{pages}{186} (\bibinfo{year}{2009}).

\end{thebibliography}
\end{document}